\documentclass [twocolumn,showpacs,preprintnumbers,amsmath,amssymb]{revtex4}

\begin{document}

\title{Probabilistic coding of quantum states}

\author{Andrzej Grudka}

\email{Andrzej.Grudka@amu.edu.pl}

\author{Antoni W\'{o}jcik}

\email{antwoj@amu.edu.pl}

\author{Miko\l aj Czechlewski}

\email{czechlewski@o2.pl}

\affiliation{Faculty of Physics, Adam Mickiewicz University,
Umultowska 85, 61-614 Pozna\'{n}, Poland}

\date{\today}

\begin{abstract}
We discuss properties of probabilistic coding of two qubits to one
qutrit and generalize the scheme to higher dimensions. We show that
the protocol preservers entanglement between qubits to be encoded
and environment and can be also applied to mixed states. We present
the protocol which enables encoding of $n$ qudits to one qudit of
dimension smaller than the Hilbert space of the original system and
then probabilistically but error-free decode any subset of $k$
qudits. We give a formula for the probability of successful
decoding.
\end{abstract}


\maketitle

\section{INTRODUCTION}

Qubit in a pure state is described by two real parameters. On the
other hand qutrit (quantum system with three-dimensional Hilbert
space) in a pure state is described by four real parameters. Thus,
the same number specifies two non-entangled qubits in a pure state
and one qutrit. Now one can ask the question if it is possible to
encode the states of these two qubits to one qutrit. Holevo bound
states that using $d$-dimensional quantum system (qudit) one can
communicate at most $\log_{2}d$ bits of classical information
\cite{Holevo}. So it is not possible to achieve the task. However,
it has been recently shown both theoretically \cite{Grudka} and
experimentally \cite{Bertuskova} that such encoding is possible in
probabilistic, but error-free manner. Namely, the protocol has been
proposed \cite{Grudka} which enables decoding with average
probability $2/3$ and perfect fidelity one arbitrarily chosen qubit
after the encoding took place. In experimental execution of the
protocol the probability of success is $1/2$. This  is due to fact
that in the experiment one replaces POVM with a von Neumann
measurement. Bertuskova \emph{et. al} \cite{Bertuskova} have shown
that the protocol can be generalized to higher dimensions.
Specifically, they have shown how to encode $N$ qudits of dimension
$d$ each in one qudit of dimension $N(d-1)+1$ and then decode one of
them. In their scheme both the procedure of encoding and decoding
succeeds with a probability strictly less than $1$. Here we present
the protocol in which the procedure of encoding is always successful
(only decoding can be unsuccessful). We also show how one can encode
$n$ non-entangled qudits to one qudit of a dimension smaller than
the Hilbert space of the original system and then probabilistically
but error-free decode any subset of $k$ qudits. A simple formula for
probability of decoding is given. We also prove that the protocol
preserves the entanglement between the qudits to be decoded and the
environment. We show that the protocol can be applied to mixed
states as well. The paper is organized as follows. In Section II we
review the original scheme and investigate its properties. In
Section III we show how one can encode two qudits of dimension $d$
in one qudit of dimension $2d-1$ and then probabilistically decode
any one of them. In Section IV we present the protocol for encoding
many qudits. We also give simple examples illustrating the protocol.
The paper ends with a brief summary in Section V.

\section{CODING OF QUBITS ENTANGLED WITH THE ENVIRONMENT AND QUBITS IN MIXED STATES}

Let us first briefly describe the original protocol. We introduce
two parties Alice and Bob. Alice performs the encoding while Bob
tries to decode the qubit with perfect fidelity. We suppose that the
states of two qubits are
\begin{equation}
|\Psi_{1}\rangle=a_{1}|0\rangle_{1}+b_{1}|1\rangle_{1}
\label{1}
\end{equation}
and
\begin{equation}
|\Psi_{2}\rangle=a_{2}|0\rangle_{2}+b_{2}|1\rangle_{2}.
\label{2}
\end{equation}
To encode the states of these qubits to one qutrit Alice performs
measurement on the joint state of the system $|\Psi_{1}\rangle
\otimes |\Psi_{2}\rangle$ given by the following measurement
operators
\begin{equation}
M_{0,0}=\frac{1}{\sqrt{3}}(|00\rangle\langle00|+|01\rangle\langle01|+|10\rangle\langle10|)
\label{3}
\end{equation}
\begin{equation}
M_{0,1}=\frac{1}{\sqrt{3}}(|01\rangle\langle01|+|00\rangle\langle00|+|11\rangle\langle11|)
\label{4}
\end{equation}
\begin{equation}
M_{1,0}=\frac{1}{\sqrt{3}}(|10\rangle\langle10|+|11\rangle\langle11|+|00\rangle\langle00|)
\label{5}
\end{equation}
\begin{equation}
M_{1,1}=\frac{1}{\sqrt{3}}(|11\rangle\langle11|+|10\rangle\langle10|+|01\rangle\langle01|).
\label{6}
\end{equation}
It should be emphasized that this is generalized measurement (POVM).
If as a result of the measurement Alice obtains $0,0$ then the state
of two qubits is projected onto a three dimensional subspace and is
now given as:
\begin{equation}
|\Psi\rangle=N(a_{1}a_{2}|00\rangle+a_{1}b_{2}|01\rangle+b_{1}a_{2}|10\rangle),
\label{7}
\end{equation}
where $N$ is normalization constant. To recover the first qubit Bob
performs a projective measurement given by the following operators:
\begin{equation}
P_{1,S}=|00\rangle\langle00|+|10\rangle\langle10|
\label{8}
\end{equation}
\begin{equation}
P_{1,F}=|01\rangle\langle01|. \label{9}
\end{equation}
If Bob obtains $1,S$ as a result of the measurement  then the state
of the qutrit is projected onto two dimensional subspace and is
identical to the state of the first qubit given by Eq. \ref{1}. If
Bob obtains $1,F$ as a result of the measurement  then the procedure
of decoding fails. Similarly to recover the second qubit Bob
performs projective measurement described by the operators:
\begin{equation}
P_{2,S}=|00\rangle\langle00|+|01\rangle\langle01| \label{10}
\end{equation}
\begin{equation}
P_{2,F}=|10\rangle\langle10|. \label{11}
\end{equation}
It should be stressed that in order to properly choose decoding
operators, Bob has to know onto which three-dimensional subspace
Alice projected the original state of two qubits. Thus Alice has to
send two bits of classical information about the result of the
measurement she obtained.

Let us now investigate the possibility of coding the entangled and
mixed states. We emphasize that the protocol does not enable
encoding of qubits in nonseparable states but this does not reject
the possibility of coding qubits entangled with two distinct
environments. Thus we suppose that each of Alice's qubits is
entangled with a qubit from the environment but they are not
correlated (neither quantum nor classically) with each other. We
assume that the first qubit and the qubit from the environment are
in the pure state:
\begin{eqnarray}
&
|\Psi_{1}\rangle=a_{1}|0\rangle_{E_{1}}|0\rangle_{1}+b_{1}|1\rangle_{E_{1}}|0\rangle_{1}+
\nonumber\\
&
c_{1}|0\rangle_{E_{1}}|1\rangle_{1}+d_{1}|1\rangle_{E_{1}}|1\rangle_{1}.
\label{12}
\end{eqnarray}
Similarly, the state of the second qubit and the qubit from the
environment is:
\begin{eqnarray}
&
|\Psi_{2}\rangle=a_{2}|0\rangle_{E_{2}}|0\rangle_{2}+b_{2}|1\rangle_{E_{2}}|0\rangle_{2}+
\nonumber\\
&
c_{2}|0\rangle_{E_{2}}|1\rangle_{2}+d_{2}|1\rangle_{E_{2}}|1\rangle_{2}.
\label{13}
\end{eqnarray}
It is convenient to write these states in the following way:
\begin{equation}
|\Psi_{1}\rangle=|\psi_{1}\rangle|0\rangle_{1}+|\phi_{1}\rangle|1\rangle_{1}
\label{14}
\end{equation}
\begin{equation}
|\Psi_{2}\rangle=|\psi_{2}\rangle|0\rangle_{2}+|\phi_{2}\rangle|1\rangle_{2},
\label{15}
\end{equation}
where
\begin{equation}
|\psi_{1(2)}\rangle=a_{1(2)}|0\rangle_{E_{1(2)}}+b_{1(2)}|1\rangle_{E_{1(2)}}
\label{16}
\end{equation}
\begin{equation}
|\phi_{1(2)}\rangle=c_{1(2)}|0\rangle_{E_{1(2)}}+d_{1(2)}|1\rangle_{E_{1(2)}}
\label{17}
\end{equation}
are in general some unnormalized and not necessarily orthogonal
vectors. If Alice performs the measurement given by operators of
Eqs. \ref{3}-\ref{6} and obtains for example $0,0$ as the result of
the measurement then the state vector of the whole system collapses
to
\begin{equation}
|\Psi\rangle=N(|\psi_{1}\rangle|\psi_{2}\rangle|00\rangle+
|\psi_{1}\rangle|\phi_{2}\rangle|01\rangle+|\phi_{1}\rangle|\psi_{2}\rangle|10\rangle).
\label{18}
\end{equation}
To recover the state of the first or the second qubit and the
corresponding qubit from the environment, Bob performs projective
measurement given by operators of Eqs. \ref{8} and \ref{9} or
\ref{10} and \ref{11} respectively. We see that the original
protocol preserves the entanglement between the qubit to be
recovered and the environment.

Let us now comment on the coding of mixed states. Let us suppose
that we have two qubits. The first (second) qubit is in a state
described by density matrix $\rho_{1(2)}$ and the state of the whole
system is:
\begin{equation}
\rho=\rho_{1}\otimes\rho_{2}. \label{19}
\end{equation}
It is well known that any mixed state can be purified
\cite{Nielsen}. Thus, we can assume that the mixed state
$\rho_{1(2)}$ is obtained from the pure state of the system and the
environment by tracing out the latter. Because the scheme preserves
entanglement between the system and the environment, the density
matrix of the qubit which is successfully decoded does not change
and we conclude that the protocol can be applied to mixed states of
the form \ref{19}.
\section{CODING OF TWO QUDITS}
Let us now describe a generalization of the scheme for coding of two
qudits. We assume that we have two non-entangled qudits of the
dimension $d$. Each of them is in a pure state (The protocol applies
to mixed states as well.):
\begin{equation}
|\Psi\rangle=\sum_{i=0}^{d-1}a_{i}|i\rangle\otimes\sum_{i=0}^{d-1}b_{i}|i\rangle.
\label{20}
\end{equation}
To encode the states of these two qudits in one qudit of dimension
$2d-1$ Alice performs the measurement described by the following
operators:
\begin{eqnarray}
& M_{i,j}=\frac{1}{\sqrt{2d-1}}(|ij\rangle\langle
ij|+\sum_{k=0,\atop k\neq i}^{d-1}|kj\rangle\langle kj|+
\nonumber\\
& +\sum_{l=0,\atop l\neq j}^{d-1}|il\rangle\langle il|). \label{21}
\end{eqnarray}
These are hermitian operators. Each term $|ij\rangle\langle ij|$ is
present in $1+2(d-1)$ operators, namely in $M_{i,j}$, $M_{k,j}$
($k\neq i$) and $M_{i,l}$ ($l\neq j$) and thus, these operators
satisfy the condition.
\begin{equation}
\sum_{i,j=0}^{d-1}M_{i,j}^{\dagger}M_{i,j}=I. \label{22}
\end{equation}
Taking it all together we see that the operators $M_{i,j}$ are
indeed the measurement operators. Each of these operators projects
the initial state of two qudits onto $(2d-1)$-dimensional subspace
of the original Hilbert space. We can now treat our system as a
qudit of the dimension $(2d-1)$. To decode the state of the first
qudit, Bob performs a projective measurement described by the
operators:
\begin{equation}
P_{1,S}=\sum_{k=0}^{d-1}|kj\rangle\langle kj| \label{23}
\end{equation}
and
\begin{equation}
P_{1,F}=\sum_{l=0,l\neq j}^{d-1}|il\rangle\langle il|. \label{24}
\end{equation}
If he obtains $1,S$ as a result of the measurement then decoding
succeeds, otherwise it fails. The procedure for decoding of the
second qudit is similar.

\section{CODING OF MANY QUDITS}

We now describe the protocol for encoding of $n$ qudits in such a
way that one can probabilistically, but error-free decode any subset
of $k$ qudits. Let us suppose that we have $n$ non-entangled qudits
of dimension $d$. Each of these qudits is in a pure state and the
state of the whole system is:
\begin{equation}
|\Psi\rangle=\sum_{i=0}^{d-1}a_{i}|i\rangle\otimes\sum_{i=0}^{d-1}b_{i}|i\rangle\otimes\sum_{i=0}^{d-1}c_{i}|i\rangle
.... \label{25}
\end{equation}
In order to encode these $n$ qudits in such a way that Bob can later
decode any subset of $k$ qudits, Alice performs a measurement
described by operators
\begin{eqnarray}
&M_{i,j,k,...}=\frac{1}{\sqrt{D_{k}}}(|ijk...\rangle\langle ijk...|+
\nonumber\\
& +\sum_{p=0,\atop p\neq i}^{d-1}|pjk...\rangle\langle
pjk...|+\sum_{q=0,\atop q\neq j}^{d-1}|iqk...\rangle\langle iqk...|+
\nonumber\\
& +\sum_{r=0,\atop r\neq k}^{d-1}|ijr...\rangle\langle ijr...|+...+
\nonumber\\
& +\sum_{p=0,q=0,\atop p\neq i,q\neq j}^{d-1}|pqk...\rangle\langle
pqk...|+\sum_{p=0,r=0,\atop p\neq i,r\neq
k}^{d-1}|pjr...\rangle\langle pjr...|+
\nonumber\\
& +\sum_{q=0,r=0,\atop q\neq j,r\neq k}^{d-1}|iqr...\rangle\langle
iqr...|+...+
\nonumber\\
& \hbox{+ OTHER TERMS}), \label{26}
\end{eqnarray}
where "OTHER TERMS" stands for similar sums over three, four ....
and $k$ indices. The constant $D_{k}$ in the above equation is equal
to the dimension of the subspace onto which $M_{i,j,k,...}$ projects
and
\begin{equation}
D_{k}=\sum_{i=0}^{k}\left( \begin{array}{ccc} n \\ i \end{array}
\right) \left( d-1 \right) ^i. \label{27}
\end{equation}
Similar arguments as before can be used to show that these operators
indeed describe a measurement. Because if $k<n$ then
\begin{eqnarray}
& D_{k}=\sum_{i=0}^{k}\left( \begin{array}{ccc} n \\ i
\end{array} \right) \left( d-1 \right) ^i <
\nonumber\\
& <\sum_{i=0}^{n}\left(
\begin{array}{ccc} n \\ i \end{array}\right) \left( d-1 \right) ^i =
\left( 1+\left( d-1\right)\right) ^n=d^n
\label{28}
\end{eqnarray}
the qudits are encoded in a system with the Hilbert space of smaller
dimension than the original one.

To decode $k$ qudits Bob performs a projective measurement described
by the operators
\begin{equation}
P_{S}= \sum_{p=0,q=0,...}^{d-1}|pqk...\rangle\langle pqk...|
\label{29}
\end{equation}
\begin{equation}
P_{F}=I-P_{S}, \label{30}
\end{equation}
where the sum is taken over indices belonging to the qudits to be
decoded and other indices are equal to those which specify the
result of the masurement \ref{26}. We will also notice that if the
qudits to be decoded are entangled between themselves then the
procedure succeeds and preserves the entanglement. However the
qudits which are decoded cannot be correlated with those which are
not decoded.

Let us now illustrate the whole protocol with a simple example. We
assume that we have three qubits. Now we can encode them in two
different ways: (1) Alice encodes three qubits in such a way that
any one of them can be later decoded and (2) Alice encodes three
qubits in such a way that any two of them can be later decoded. In
both cases the initial state of the system is:
\begin{eqnarray}
&
|\Psi\rangle=a_{1}a_{2}a_{3}|000\rangle+a_{1}a_{2}b_{3}|001\rangle+
\nonumber\\
&
+a_{1}b_{2}a_{3}|010\rangle+a_{1}b_{2}b_{3}|011\rangle+b_{1}a_{2}a_{3}|100\rangle+
\nonumber\\
& +b_{1}a_{2}b_{3}|101\rangle+
b_{1}b_{2}a_{3}|110\rangle+b_{1}b_{2}b_{3}|111\rangle. \label{31}
\end{eqnarray}

In the case of the first coding we have $n=3$, $d=2$ and $k=1$. To
encode three qubits Alice projects the state of the system on
$4$-dimensional subspace with measurement operators defined in Eq.
\ref{26}, for example
\begin{eqnarray}
& M_{0,0,0}=\frac{1}{2}(|000\rangle\langle 000|+|001\rangle\langle
001|+
\nonumber\\
& +|010\rangle\langle 010|+|100\rangle\langle 100|). \label{32}
\end{eqnarray}
If Alice obtains $0,0,0$ as a result of her measurement then the
state of the system becomes
\begin{eqnarray}
& |\Psi\rangle=N(a_{1}a_{2}a_{3}|000\rangle+
a_{1}a_{2}b_{3}|001\rangle+
\nonumber\\
& +a_{1}b_{2}a_{3}|010\rangle+b_{1}a_{2}a_{3}|100\rangle).
\label{33}
\end{eqnarray}
If Bob wants to decode the state of the first qubit he performs a
projective measurement described by the operators
\begin{equation}
P_{1,S}=|000\rangle\langle 000|+|100\rangle\langle 100| \label{34}
\end{equation}
\begin{equation}
P_{1,F}=|010\rangle\langle 010|+|001\rangle\langle 001|. \label{35}
\end{equation}
If he obtains $1,S$ as a result of the measurement then he
successfully decodes the first qubit.

In the case of the second coding we have $n=3$, $d=2$ and $k=2$. To
encode three qubits Alice projects the state of the system onto a
$7$-dimensional subspace with measurement operators defined in Eq.
\ref{26}, for example
\begin{eqnarray}
& M_{0,0,0}=\frac{1}{\sqrt{7}}(|000\rangle\langle 000|+
\nonumber\\
& +|001\rangle\langle 001|+|010\rangle\langle
010|+|100\rangle\langle 100|+
\nonumber\\
& +|011\rangle\langle 011|+|101\rangle\langle
101|+|110\rangle\langle 110|). \label{36}
\end{eqnarray}
It should be noted here that the dimension of the space onto which
the original space is projected depends on two things: (1) the
number of qudits to be encoded and (2) the number of qudits to be
decoded.

If Alice obtains $0,0,0$ as a result of her measurement then the
state of the system becomes
\begin{eqnarray}
&
|\Psi\rangle=N(a_{1}a_{2}a_{3}|000\rangle+a_{1}a_{2}b_{3}|001\rangle+
\nonumber\\
& a_{1}b_{2}a_{3}|010\rangle+b_{1}a_{2}a_{3}|100\rangle+
a_{1}b_{2}b_{3}|011\rangle+
\nonumber\\
& b_{1}a_{2}b_{3}|101\rangle+ b_{1}b_{2}a_{3}|110\rangle).
\label{37}
\end{eqnarray}
If Bob wants to decode the state of the first and the second qubit
he performs a projective measurement described by the operators
\begin{eqnarray}
& P_{S}=|000\rangle\langle 000|+|100\rangle\langle 100|+|010\rangle
\nonumber\\
& \langle 010|+|110\rangle \langle 110|
\label{38}
\end{eqnarray}
\begin{equation}
P_{F}=|001\rangle\langle 001|+|011\rangle\langle
011|+|101\rangle\langle 101|. \label{39}
\end{equation}
If he obtains $S$ as a result of the measurement then he has
successfully decoded the first and the second qubit.

In the protocols described the procedure of decoding is always
successful, however one does not know onto which subspace the
initial state of the system will be projected. The choice of the
decoding measurement depends on (1) the qudits to be decoded and (2)
the subspace onto which the initial state was projected. Because of
the latter Alice must send to Bob $k\log_{2} d$ bits of classical
information about the result of her measurement. The procedure of
decoding is probabilistic and it only succeeds with some
probability. If we assume that each qudit is prepared in a randomly
chosen pure state then the average probability of successful
decoding of $k$ qudits of $n$ encoded qudits is equal to:
\begin{equation}
p_{S}=\frac{d^{k}}{D_{k}}.
 \label{40}
\end{equation}
This is the dimension of the Hilbert space of $k$ decoded qudits
divided by the dimension of the Hilbert space of the qudit to which
$n$ qudits were encoded.

\section{SUMMARY}

We have shown that scheme of Ref. \cite{Grudka} preserves the
entanglement between the qubit to be decoded and the environment and
can be also used for coding of mixed states. We also presented a
much more general protocol which enables encoding of $n$ qudits in a
one qudit of dimension smaller than the dimension of the Hilbert
space of the original system and then probabilistically decode any
subset of $k$ of them. The probability of successful decoding is
equal to the dimension of the Hilbert space of $k$ qudits divided by
the dimension of the qudit in which $n$ qudits are encoded.

\begin{acknowledgments}
Two of us (A.G. and A.W.) would like to thank the State Committee
for Scientific Research for financial support under Grant No. 0 T00A
003 23.
\end{acknowledgments}

\end{document}